\documentclass[prd,aps]{revtex4}

\usepackage{epsfig}

\newcommand{\be}{\begin{equation}}
\newcommand{\ee}{\end{equation}}
\newcommand{\bea}{\begin{eqnarray}}
\newcommand{\eea}{\end{eqnarray}}

\begin{document}

\title{
Production of the charged Higgs bosons at the CERN Large Hadron Collider 
\\
in the left-right symmetric model
}

\author{ Dong-Won Jung }
\email{dwjung@ncu.edu.tw}

\affiliation{
Department of Physics, National Central University, 
Jhong-Li, Taiwan, 32054.
}

\author{ Kang Young Lee }
\email{kylee@muon.kaist.ac.kr}

\affiliation{
Department of Physics, Korea University,
Seoul 136-713, Korea. 
}

\date{\today}

\begin{abstract}

We study the production of the charged Higgs boson
at the LHC in the left-right symmetric model.
It is shown that there exists a lower bound of the cross section.
We investigate that predicted cross sections of this model are 
generally larger than those of the two Higgs doublet model 
or the minimal supersymmetric model.

\end{abstract}

\maketitle

\section{Introduction}

Search of Higgs bosons is the primary goal 
of the CERN Large Hadron Collider (LHC). 
In the standard model (SM), 
one neutral Higgs boson exists as a result 
of the electroweak symmetry breaking (EWSB),
of which mass is not predicted in the theoretical framework.
The discovery potential for the SM Higgs boson at the LHC 
will reach to about 1 TeV with the integrated luminosity
of 100 fb$^{-1}$.
It implies that the SM Higgs boson will be discovered
at the LHC at more than 5-$\sigma$ level.
In many models of new physics beyond the SM, however,
more symmetries are involved and 
the Higgs sector should be extended to break larger symmetry.
Signatures of an extended Higgs sector would provide 
direct evidence for new physics beyond the SM.
Models with an extended Higgs sector often contain
charged Higgs bosons $H^\pm$, which do not exist within the SM.

The two Higgs doublet model (2HD) is one of the simplest
extension of the SM and the Higgs structure of
the minimal supersymmetric standard model (MSSM).
There are three neutral Higgs bosons, $h^0$, $H^0$, $A^0$ 
and a pair of charged Higgs bosons $H^\pm$
in the 2HD model and the MSSM.
Hereafter the 2HD model and the MSSM 
are just called the 2HD model generically.
If the $H^\pm$ bosons are lighter than the top quark,
they can be produced in the top quark decay processes 
$t \to b H^{+}$.
Search for $H^\pm$ from top quark decays
in the 2HD model has been performed at Tevatron 
and no evidence for $H^\pm$ production is found \cite{cdf}.
The CERN Large Electron Positron Collider (LEP) 
has also examined the charged Higgs bosons
up to $\sqrt{s} = 200$ GeV
through the pair production of $H^\pm$ 
to present the lower bound of the charged Higgs boson mass 
\cite{lep,borzumati}.
Recent measurement of Br($B^\pm \to \tau \nu$) by Belle
provides indirect constraints on the charged Higgs bosons
in the 2HD model
via the annihilation diagram mediated by $H^\pm$ boson 
\cite{btaunu,hou}.
The absence of the observed charged Higgs boson so far
derives constraints on $(\tan \beta, m_{H^\pm})$ parameter space
for the 2HD model.
If the charged Higgs bosons are heavier than the top quark,
we have to observe the direct production at hadron colliders.
The most promising channel for the charged Higgs boson 
production at the LHC is the $gb \to tH^\pm$ process 
which has been extensively studied
\cite{gbth1,gbth2,gbth3,belyaev}.
The Drell-Yan mechanism $gg$, $q \bar{q} \to H^- H^+$
and the associated production with a $W$ boson,
$q \bar{q} \to H^\pm W^\mp$ are suppressed
due to the weak couplings, low quark luminosity, and loop suppression.
The discovery potential of the $H^\pm$ bosons at the LHC
has been studied by ATLAS \cite{atlas} and CMS \cite{cms} groups.
Expected to be discovered is a MSSM charged Higgs boson 
as heavy as 1 TeV at the 5-$\sigma$ confidence level,
or may be excluded up to the mass of 1.5 TeV at 95 \% C.L. at the LHC
with the MSSM radiative corrections \cite{belyaev}.

The left-right (LR) symmetric model based on the gauge symmetry,
SU(2)$_L \times$ SU(2)$_R \times$ U(1)$_{B-L}$,
usually contain a bidoublet Higgs field $\phi (2, \bar{2}, 0)$ 
for the EWSB and the Yukawa couplings
represented by
\be
\phi =
\left(
\begin{array}{cc}
\phi_1^0 & \phi_1^+ \\
\phi_2^- & \phi_2^0 \\
\end{array}
\right).
\ee
since both of the right-handed fermions and the left-handed fermions 
transform as doublets under SU(2)$_R$ and SU(2)$_L$
\cite{LR}.
The additional SU(2)$_R$ gauge symmetry should be broken
by another Higgs sector
at the scale much higher than the electroweak scale
\cite{czagon,cheung,chay,erler}.
The weak scale phenomenology of the Higgs sector 
is principally determined by the bidoublet Higgs fields,
and the dominant field contents are similar to those of the 2HD model:
$h^0$, $H^0$, $A^0$ and $H^\pm$.
However, the structure of the Yukawa couplings and potential 
of the bidoublet Higgs fields are much different from
those of the two doublet Higgs fields.
Therefore, it leads to different charged Higgs boson phenomenologies
from those in the 2HD model.
Constraints on the charged Higgs boson parameter space in the LR model
with the present experimental data from Tevatron, LEP and Belle 
has been presented in Ref.
\cite{chargedhiggs}.
In this work, we explore the production of $H^\pm$ bosons
at the LHC in the LR model. 
This paper is organized as follows:
In section 2, we briefly review the Higgs sector of the LR model.
The production cross sections at the LHC 
are presented in section 3 and we conclude in section 4.

\section{The Higgs sector of the left-right symmetric model}

The left-right symmetric model involves an additional SU(2)$_R$ symmetry 
which has to be broken at the higher scale than the electroweak scale.
Two triplet Higgs fields 
$\Delta_L (3,1,2)$ and $\Delta_R (1,3,2)$ represented by
\be
\Delta_{L,R} =
\frac{1}{\sqrt{2}}
\left(
\begin{array}{cc}
\delta^+_{L,R} & \sqrt{2} \delta^{++}_{L,R}  \\
\sqrt{2} \delta^0_{L,R} & -\delta^{+}_{L,R}  \\
\end{array}
\right) ,
\ee
are introduced to break the additional symmetry of the model.
Actually $\Delta_R$ breaks the SU(2)$_R$ symmetry and 
another triplet $\Delta_L$ is just introduced
as a result of manifest left-right symmetry.
The kinetic terms for Higgs fields are given by 
\be 
{\cal L} = {\bf Tr} \left[ ( D_\mu \Delta_{L,R} )^\dagger 
                       ( D^\mu \Delta_{L,R} )\right] 
               + {\bf Tr} \left[ (D_\mu \phi)^\dagger 
                        (D^\mu \phi) \right], 
\ee 
where the covariant derivatives are defined by 
\bea 
D_\mu \phi &=& \partial_\mu \phi - i \frac{g}{2} W^a_{L \mu} \tau^a \phi 
              + i\frac{g}{2} \phi W^a_{R \mu} \tau^a , 
\nonumber\\
D_\mu \Delta_{L,R} &=& \partial_\mu \Delta_{L,R}
          - i\frac{g}{2} \left[W^a_{L,R \mu} \tau^a , \Delta_{L,R} \right]
          - i g^\prime B_\mu \Delta_{L,R}.
\eea
The spontaneous breaking of gauge symmetries is triggered
by the vacuum expectation values (VEV)
\be
\langle \phi \rangle = \frac{1}{\sqrt{2}}\left(
\begin{array}{cc}
  k_1&0 \\
  0&k_2 \\
\end{array}
\right),
~~~~~~~~~~~~
\langle \Delta_{L,R} \rangle = \frac{1}{\sqrt{2}}\left(
\begin{array}{cc}
  0&0 \\
  v_{L,R}&0 \\
\end{array}
\right).
\ee
The SU(2)$_{\rm R}$ breaking scale $v_R$ 
should be higher than the electroweak scale, 
$k_{1,2} \ll v_R$ since $W_R$ should be heavier than $W_L$.  
Note that $v_L$ is irrelevant for the symmetry breaking 
and just introduced in order to manifest the left-right symmetry.  
The see-saw relation for the neutrino mass 
$m_\nu \sim M_{LL}+M^2_{LR}/M_{RR}$ 
tells us that
$v_R$ is typically very large $\sim 10^{11}$ GeV,
where $M_{ij}$ are the matrix elements for the masses
in the $(W_L,W_R)$ basis.  
Then the heavy gauge bosons are too heavy to be produced
at the accelerator experiments
and the SU(2)$_R$ structure is hardly probed in the laboratory.
Provided that $v_R$ is assumed to be only moderately large 
$v_R \sim {\cal O} ({\rm TeV})$
for the heavy gauge bosons to be studied at LHC,
the Yukawa couplings require to be suppressed
in order that the neutrino masses are at the eV scale
and $v_L$ should be very small or close to 0.
This is achieved when the quartic couplings of
$(\phi \phi \Delta_L \Delta_R)$-type terms in the Higgs potential
are set to be zero \cite{gunion2,kiers}.
This limit is warranted 
by the approximate horizontal U(1) symmetry \cite{khasanov}
as well as the see-saw picture for light neutrino masses.  
We adopt this limit here.
Higgs boson masses are not affected by taking this limit \cite{kiers}.

The general Higgs potential in the LR model has been studied in Refs. 
\cite{gunion1,gunion2,kiers}.
We define the parameters $\xi = k_2 / k_1$ and
$\epsilon = k_1 / v_R $ for convenience. 
The parameter $\xi$ is the ratio of two VEVs for the EWSB
which is corresponding to $\tan \beta$ in the 2HD model.
Since $\epsilon \ll 1$, we will present our formulas as powers of $\epsilon$.
Taking the limit that the quartic couplings
for $(\phi \phi \Delta_L \Delta_R)$ terms and $v_L$ go to 0
as mentioned above,
the charged Higgs boson mass matrix is given
in the basis of $(  \phi_1^+ ,\phi_2^+ ,\delta_R^+ ,\delta_L^+ )$ by
\be
{\cal M}^2_+ = \left(%
\begin{array}{ccccccc}
m_+^2     && m_+^2 \xi   && m_+^2 \epsilon (1-\xi^2)/\sqrt{2}     && 0\\
m_+^2 \xi && m_+^2 \xi^2 && m_+^2 \epsilon \xi (1-\xi^2)/\sqrt{2} && 0 \\
m_+^2 \epsilon (1-\xi^2)/\sqrt{2} && m_+^2 \epsilon \xi (1-\xi^2)/\sqrt{2}&& 
                          m_+^2 \epsilon^2 (1-\xi^2)^2/2        && 0 \\
0         && 0           && 0                    && m^{(+)^2}_{\rho_3} \\
\end{array}%
\right), 
\ee
where $ m_+^2 = \alpha_3 v_R^2/2 (1-\xi^2)$
with the quartic coupling $\alpha_3$ for
${\rm Tr}(\phi^\dagger \phi \Delta_L \Delta_L^\dagger)
+{\rm Tr}(\phi^\dagger \phi \Delta_R \Delta_R^\dagger)$ term
\cite{kiers}.
If $\xi >1$, $\alpha_3$ should be negative to avoid 
the dangerous negative mass square of scalar fields.
Note that $\delta_L^+$ field from the Higgs triplet $\Delta_L$
decouples from other three charged Higgs fields
with mass $ m^{(+)^2}_{\rho_3}$
and is irrelevant for our phenomenological discussion here.
The mass of the lightest charged Higgs boson is given by
\be
m^2_{H^\pm} = m_+^2 (1+\xi^2)
     \left( 1+ \frac{1}{2} \epsilon^2 \frac{(1-\xi^2)^2}{1+\xi^2} \right) ,
\ee
after diagonalization.
Since $m_{H^\pm}$ depends on the coupling $\alpha_3$,
it is an independent observable.
Thus the charged Higgs phenomenologies are
expressed in terms of two parameters
$\xi$ and $m_{H^\pm}$.
Note that the mass of $W_R$ boson does not appear
at the charged Higgs phenomenology.
No CP violation in the Higgs sector is assumed for simplicity. 

Violating the lepton numbers and baryon numbers,
the triplet Higgs fields do not allow 
the ordinary Yukawa coupling terms for Dirac fermions.
Thus quark and lepton masses are derived from the Yukawa couplings 
in terms of the bidoublet Higgs fields,
given by
\bea
{\cal L} = {\bar \Psi^i_L} \left( {\cal F}_{ij} \phi 
           + {\cal G}_{ij} {\tilde \phi} \right) \Psi^j_R + H.c.,
\eea
where $ \Psi^i = ( {\hat U}, {\hat D} )^\dagger$ is
the flavour eigenstates, $ {\tilde \phi} = \tau_2 \phi^\ast \tau_2$,
and ${\cal F}$, ${\cal G}$ are $3\times3$ Yukawa coupling matrices.
We rotate $ {\hat U}$ and $ {\hat D} $ into the mass eigenstates
by unitary transforms,
${\hat U}_{L,R} = V^U_{L,R} U_{L,R}$ and
${\hat D}_{L,R} = V^D_{L,R} D_{L,R}$,
to define Cabibbo-Kobayashi-Maskawa (CKM) matrix
$V^{CKM}_{L,R} = {V^U_{L,R}}^\dagger V^D_{L,R}$.
We assume the manifest left-right symmetry 
$V^{CKM}_L= V^{CKM}_R$.
The Yukawa coupling matrices ${\cal F}$ and ${\cal G}$
are given in terms of 
\bea
{\cal F} &=& \frac{\sqrt{2}}{k_-^2}
         \left(  k_1 V^U_L {\cal M}^U {V^U_R}^\dagger
            -k_2 V^D_L {\cal M}^D {V^D_R}^\dagger \right),
\nonumber \\
{\cal G} &=& \frac{\sqrt{2}}{k_-^2}
        \left( -k_2 V^U_L {\cal M}^U {V^U_R}^\dagger
            +k_1 V^D_L {\cal M}^D {V^D_R}^\dagger \right),
\eea
where $k_-^2 = |k_1|^2 - |k_2|^2$ and
${\cal M}^U$ and ${\cal M}^D$ are diagonal mass matrices
for $U$-type and $D$-type quarks respectively.
Note that these solutions for the Yukawa coupling matrices 
no longer hold for $\xi=1$ we have to treat the $\xi=1$ case 
in a separate way.
We do not consider that case in this work.
Although small $\xi$ is preferred in order to generate the ratio $m_b/m_t$, 
$\xi >1$ region cannot be excluded in general.

\section{production of the charged Higgs boson at the LHC}

The relevant interaction lagrangian of
the charged Higgs boson production is given by
\be
-{\cal L} = V_{tb}^{\ast}~ \bar{b} ( g_L P_L + g_R P_R) t H^- + {\rm H.c.},
\ee
where the couplings are defined by
\bea g_L &=& \sqrt{ 2 \sqrt{2} G_F} 
         \left( m_U \frac{1+\xi^2}{|1-\xi^2|}
               - m_D \frac{2 \xi}{|1-\xi^2|} \right)
         \left( 1 - \frac{1}{4} \epsilon^2 (1+\xi^2) \right)
     + {\cal O}(\epsilon^4),
\nonumber \\
g_R &=& \sqrt{ 2 \sqrt{2} G_F} 
         \left( m_U \frac{2 \xi}{|1-\xi^2|}
               - m_D \frac{1+\xi^2}{|1-\xi^2|} \right)
         \left( 1 - \frac{1}{4} \epsilon^2 (1+\xi^2) \right)
     + {\cal O}(\epsilon^4).
\eea

\begin{center}
\begin{figure}[t]
\hbox to\textwidth{\hss\epsfig{file=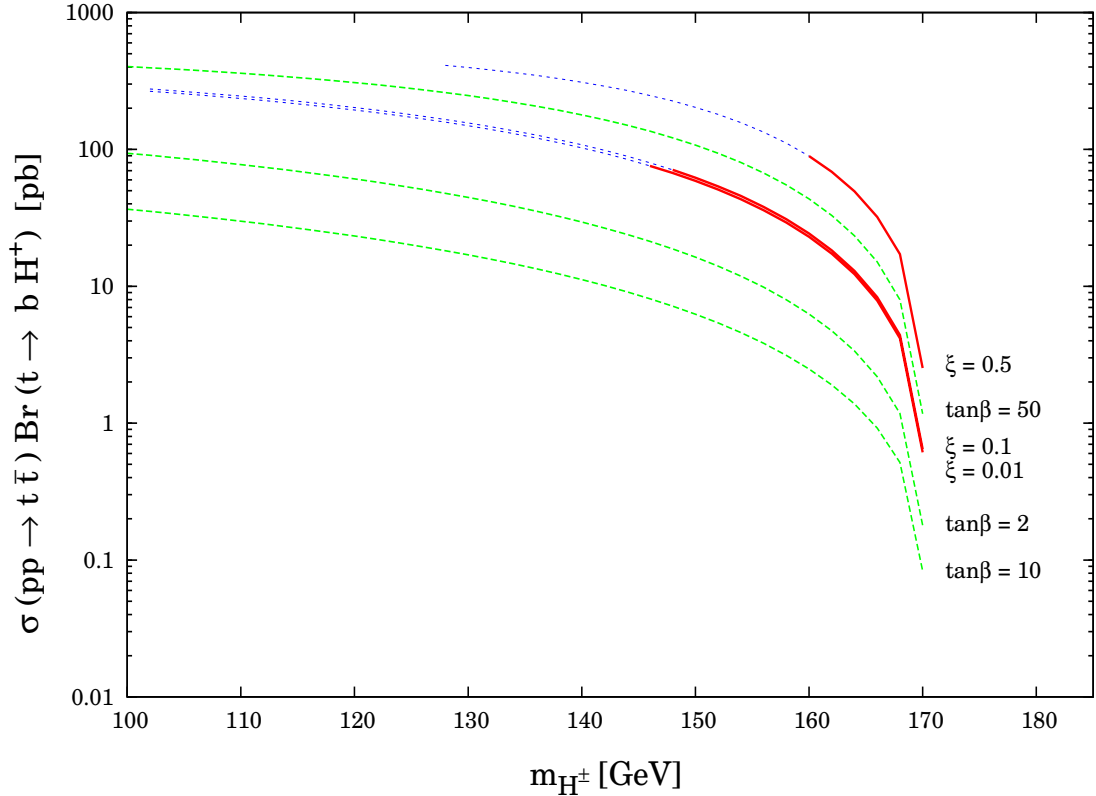,width=15cm,height=11cm}\hss}
 \vskip -1.5cm
\vspace{1cm}
\caption{
Production cross sections of a light charged Higgs boson through
the sequential decay $p p \to t \bar{t} \to \bar{t} b H^+$ at the LHC
with respect to $H^\pm$ masses
when $m_{H^\pm} < m_t-m_b$.
}
\end{figure}
\end{center}

The light charged Higgs boson such that $m_{H^\pm} < m_t-m_b$
can be produced through the top quark decay $t \to b H^\pm$ sequentially
after the top quark pair production at the LHC.
The cross section of top quark pair production 
$\sigma ( p p \to t \bar{t}) = {\cal O} (1 ~{\rm nb})$
implies that $10^8$ pairs of top quarks will be produced with
 expected integrated luminosity of 100 fb$^{-1}$ at the LHC.
We show the light charged Higgs boson production
defined by the cross section times branching ratio,
$\sigma(pp \to t \bar{t}) \cdot {\rm Br}(t \to b H^+)$
in Fig. 1.
We use the total cross section for $t \bar{t}$ production
as 833 pb for next-to-leading order (NLO) QCD corrections including
next-to-leading-log (NLL) resummation \cite{NLONLL}.
The branching ratio of the top quark decay is constrained 
by the measurement at the Tevatron, 
${\rm Br}(t \to Wb) = 0.94^{+0.31}_{-0.24}$
\cite{pdb}.
The decay of the top quark into a light $H^\pm$ boson
has been examined by the CDF collaboration at Tevatron
to obtain the exclusion region on the model parameter space
with the absence of the  observed charged Higgs boson
in the MSSM \cite{cdf} and in the LR model \cite{chargedhiggs}.
The parameter space $(\xi, m_{H^\pm})$ of the LR model 
is severely constrained by the Tevatron data, 
even there is a conservative lower bound of $m_{H^\pm} > 145$ GeV.
The solid parts of the plots are predictions with allowed parameters
in the LR model.
As a benchmark, the productions of $H^\pm$ in the MSSM are also plotted
with varying $\tan \beta$.
Yukawa couplings of the LR model have a definite lower bound 
as $\xi$ varies, so do the production cross sections.
We may consider the graph for $\xi = 0.01$ to be the lower bound
since the Yukawa couplings are saturated as $xi$ becomes smaller.
The cross section with $\xi=0.01$ is almost same as 
that of the 2HD model with $\tan \beta \approx 35$.
Thus we find that the productions in the LR model are generically larger than
those in the 2HD model unless $\tan \beta$ is large.

\begin{center}
\begin{figure}[t]
\hbox to\textwidth{\hss\epsfig{file=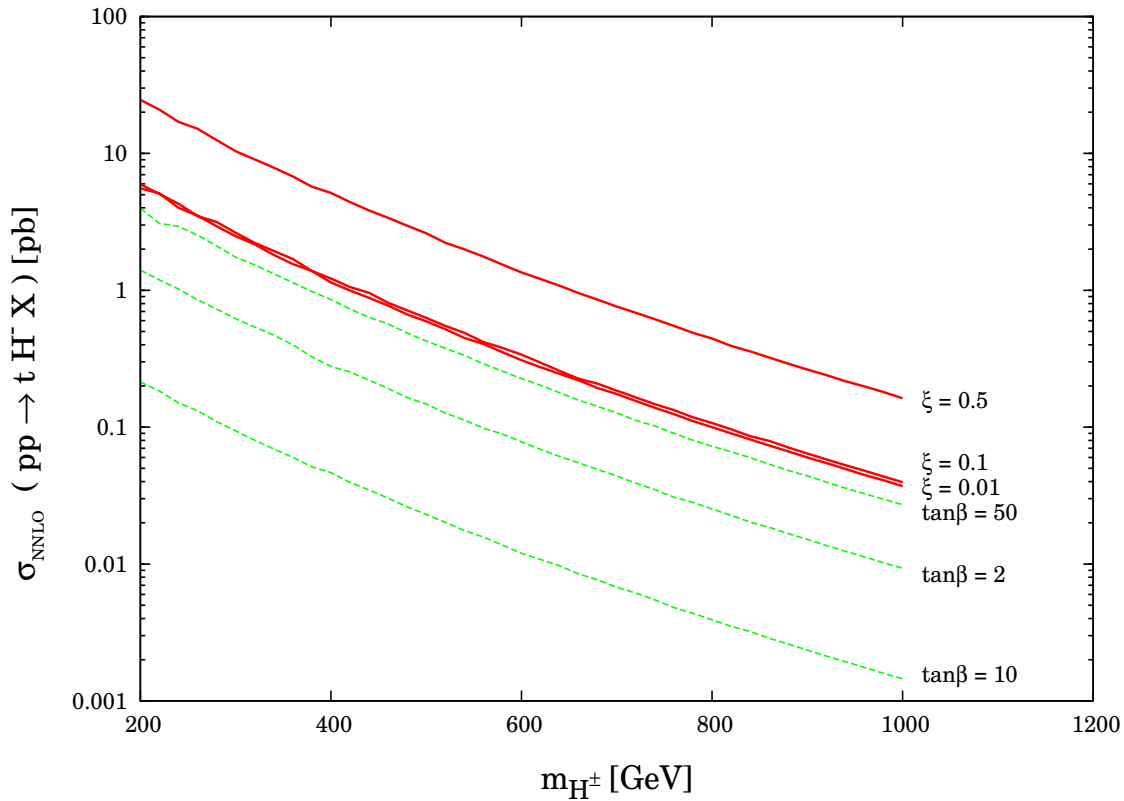,width=15cm,height=11cm}\hss}
 \vskip -1.5cm
\vspace{1cm}
\caption{
Cross sections of $p p \to g b \to \bar{t} H^+$ process at the LHC
including NNLO QCD corrections
with respect to the charged Higgs boson masses. 
}
\end{figure}
\end{center}

\begin{center}
\begin{figure}[ht]
\hbox to\textwidth{\hss\epsfig{file=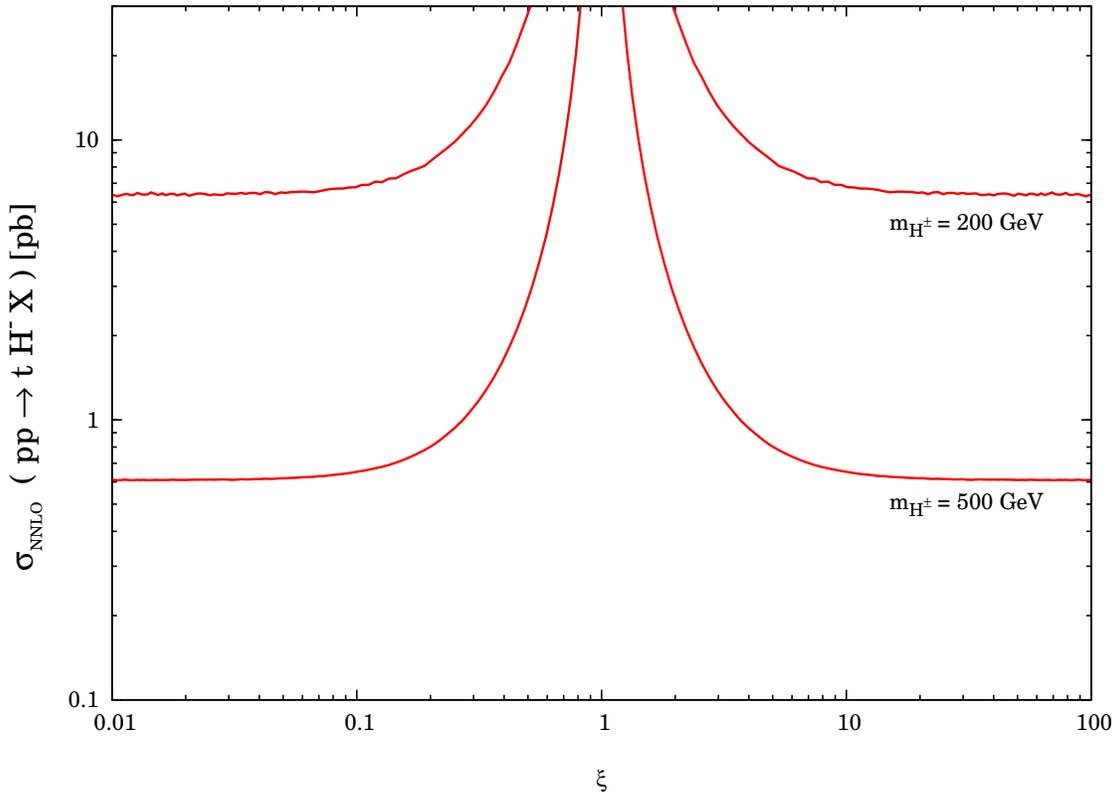,width=15cm,height=11cm}\hss}
 \vskip -1.5cm
\vspace{1cm}
\caption{
Cross sections of $p p \to g b \to \bar{t} H^+$ process at the LHC
including NNLO QCD corrections
with respect to $\xi$.
}
\end{figure}
\end{center}
If $H^\pm$ is heavier than the top quark, 
the $gb \to tH^\pm$ process is the most promising channel 
for direct production of $H^\pm$ boson at the LHC.
We write the scattering amplitude as
\be
i{\cal M}(g \bar b \to \bar t H^+)
= i g_s \bar{b} \left[ T_{ab}^c \not{\epsilon}(p_g)
          \frac{g_L P_L + g_R P_R}{\not{p}_H - \not{p}_b - m_t}
        + \frac{g_L P_L + g_R P_R}{\not{p}_H + \not{p}_t - m_b}
            \not{\epsilon}(p_g) T_{ab}^c
        \right]t,
\ee
in terms of the Yukawa couplings given in Eq. (11)
and obtain the partonic cross section as
\be
\hat \sigma(g \bar{b} \to \bar{t} H^+) =
\frac{1}{32 \pi \hat s}
\frac{\sqrt{(\hat s - m_t^2 - m_H^2)^2 -4 m_H^2 m_t^2}}{\hat s - m_b^2}
|\bar {\cal M} |^2,
\ee
where the colliding energy is kinematically allowed
if $\hat s > (m_t + m_H)^2$.
The hadronic cross section is given by
\be
\sigma(p p \to g \bar{b} \to \bar{t} H^+)
       = \int dx~dy~f_g(x)~f_b(y)~\hat \sigma(g \bar{b} \to \bar{t} H^+),
\ee
where $f_{g}$ and $f_b$ are the parton distribution functions (PDF)
for gluon and $b$-quark respectively.
We use the leading order CTEQ6 functions for a gluon and a $b$-quark
PDF in a proton \cite{CTEQ6}.
The QCD factorization and renormalization scales $Q$ are set to be the
$g b$ invariant mass,  {\it i.e.}, $\sqrt{\hat{s}}$.
The NLO QCD corrections 
to $g b \to t H^-$ process in the MSSM
has been calculated in Refs. \cite{NLO,NLOSUSY} 
and the next-to-next-to-leading order (NNLO)
soft-gluon corrections calculated in Ref. \cite{NNLO}.
They are shown to be substantial contributions 
to the production cross section and we include them.
Since the QCD corrections are model-independent,
we can use the $K$-factors for the cross section 
given in Ref. \cite{NNLO} 
in order to include the QCD corrections for our work. 
The cross sections including NNLO corrections are depicted in Fig. 2.
As a benchmark, the cross sections in the MSSM are also plotted.
The running masses for top quark and bottom quark 
at the scale of $ m_{H^\pm}$
have been used in the formula, e.g. $m_t \approx 170$ GeV
and $m_b \approx 2.9$ Gev for $m_{H^\pm} = 200$ GeV.
The supersymmetric NLO QCD corrections calculated in Ref. \cite{NLOSUSY}
are relatively small and depends on the SUSY parameters
and we do not include them here.
Contribution from the $2 \to 3$ process, $gg \to t \bar{b} H^-$
is not considered in this paper.
This is considerable, but essentially common process
which shows the same trends with a factor two or three smaller 
than the  $g b \to t H^-$ process in both models.
The $\xi$--dependence of the cross section is shown in Fig. 3.
Near $\xi = 1$, the cross section drastically increases
and it is saturated as $\xi$ goes far off.

As in the case of the light $H^\pm$,
there is a lower bound for the production cross section 
$\sigma(pp \to gb \to \bar{t} H^+)$ in the LR model.
The lower bound value of the cross section of the LR model is 
close to that of the 2HD model with $\tan \beta=57$.
Thus the cross section of the LR model is generally larger than 
that in the 2HD model except for large $\tan \beta$ region.
We estimate that 
\be
\frac{\left( {g_L}^2 + {g_R}^2 \right)_{LR}}
     {\left( {g_L}^2 + {g_R}^2 \right)_{2HD}}
\approx
\frac{m_t^2 \left( (1+\xi^2)^2/(1-\xi^2)^2 + 4 \xi^2/(1-\xi^2)^2 \right)} 
{m_b^2 \tan^2 \beta  + m_t^2 \cot^2 \beta}
\gtrsim 2,
\ee
when $\tan \beta \lesssim 50$.
The cross section in the 2HD model increases as $\tan \beta$ increases
but suppressed by $m_b^2$ and the $m_t^2$ term is suppressed
by $1/\tan^2 \beta$. 
In the LR model, the Yukawa couplings involve the factors of
$2 \xi/|1-\xi^2|$ or $(1+\xi^2)/|1-\xi^2|$ instead of
$\tan \beta$ or $1/\tan \beta$, and they are of order ${\cal O}(1)$
or more.
Thus the $m_t^2$ terms in the cross section dominate and 
the cross section is larger than that in the 2HD model
unless $\tan \beta$ is large enough.

\section{Decay of the charged Higgs boson}

The produced charged Higgs boson will decay into light particles.
The dominant decay channel is $H^+ \to t \bar{b}$
for $m_{H^\pm} > m_t + m_b$
due to large top quark mass.
Another interesting channel is $H^+ \to \bar{\tau} \nu$
by triggering with a high $p_T$ lepton.
We consider the ratio of the decay width for $H^+ \to \bar{\tau} \nu$
to that for $H^+ \to t \bar{b}$ in the LR model and 2HD model.
In the LR model, we have
\be
\frac{\Gamma(H^+ \to \bar{\tau} \nu)}
     {\Gamma(H^+ \to t \bar{b})}
     = \frac{1}{3 |V_{tb}|^2 \left( 1-m_t^2/m_{H^\pm}^2 \right)}
      \frac{m_\tau^2}{m_t^2}
       \frac{4 \xi^2}
            {1+6 \xi^2 + \xi^4}, 
\ee
while the ratio in the 2HD model is given by
\be
\frac{\Gamma(H^+ \to \bar{\tau} \nu)}
     {\Gamma(H^+ \to t \bar{b})}
     = \frac{1}{3 |V_{tb}|^2 \left( 1-m_t^2/m_{H^\pm}^2 \right)}
       \frac{m_\tau^2 \tan^2 \beta}
            {m_t^2 \cot^2 \beta + m_b^2 \tan^2 \beta},
\ee
where $m_b/m_{H^\pm},~m_\tau/m_{H^\pm} \ll 1$
and QCD corrections are ignored. 
The ratios are depicted in Fig. 4.
We find that the ratio can be sizable in the 2HD model 
as $\tan \beta$ increases.
It is as large as more than 10 $\%$ when $\tan \beta > 10$,
while the ratio is always negligible in the LR model
by suppression of $m_\tau^2/m_t^2$.
Therefore we can discriminate the underlying Higgs structure
of the charged Higgs boson by measuring the ratio
$\Gamma(H^+ \to \bar{\tau} \nu)/\Gamma(H^+ \to t \bar{b})$
in the most interesting region of parameter space.
When $m_{H^\pm} < m_t - m_b$, 
the tau channel dominates like in the 2HD model.

\begin{center}
\begin{figure}[ht]
\hbox to\textwidth{\hss\epsfig{file=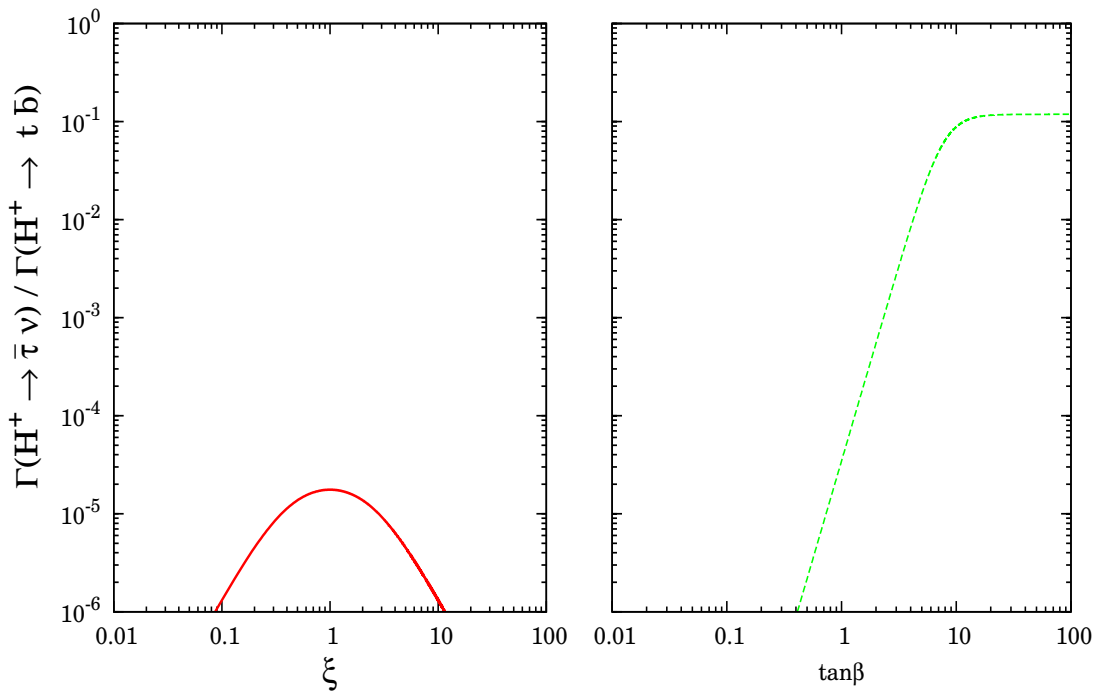,width=15cm,height=11cm}\hss}
 \vskip -1.5cm
\vspace{1cm}
\caption{The ratios
$\Gamma(H^+ \to \bar{\tau} \nu)/\Gamma(H^+ \to t \bar{b})$
with respect to $\xi$ and $\tan \beta$ in the LR model and the 2HD model
respectively.
} 
\end{figure}
\end{center}

\section{Detection of the charged Higgs boson}

As shown in the previous section, 
the $H^\pm$ boson in the LR model
mostly decays into $t \bar{b} (\bar{t} b)$
when $m_{H^\pm} > m_t + m_b$.
The decay product contains two top quarks.
We require one of the top quark to decay semileptonically,
$ t \to W b \to l \nu b$,
and the other hadronically, $t \to W b \to j j' b$,
for triggering and reconstruction of the top quark as well as 
the charged Higgs boson.
The promising final states inside the detector are given by
\be 
g b \to t H^\pm \to t t b \to W^+ W^- b b b \to l \nu j j' b b b,
\ee
which involve three $b$-jets and an isolated lepton.
The main background in the SM is coming from 
the $t \bar{t}$ pair production involving another $b$ jet
or other jets.
The Monte Carlo simulations for the detection of the charged Higgs boson
in the MSSM through $H^\pm \to t b$ have been performed by
ATLAS \cite{atlasHtb} and CMS \cite{cmsHtb} groups 
to estimate the discovery potentials of $H^\pm$.
The essential improvement is expected in the signal reconstruction
if the $b$-tagging efficiency increases in the future.
Their results show that the charged Higgs can be discovered
up to 300 GeV 
at low luminosity period of the LHC with the 
integrated luminosity $\int {\cal L} \sim 30$ fb$^{-1}$.
The discovery limit can be raised to be 400 GeV 
for high $\tan \beta$ ($> 25$).
It is hard to discover the charged Higgs of the 2HD model 
with the mass above 400 GeV.
On the other hand, 
the production cross sections of $H^\pm$ in the LR model 
are generically larger than those in the 2HD model.
The larger number of produced $H^\pm$ bosons 
due to the larger production cross section 
indicates an improvement of the signal-to-background ratio
since the background events are the SM processes.
We expect the better visibility of the charged Higgs boson
in the LR model, although no Monte Carlo study for $H^\pm$
has been done yet in the LR model.

\section{Concluding Remarks}

In this work, we have calculated the production cross sections of 
the charged Higgs boson in the LR model at the LHC.
Observation of a charged Higgs boson is a clear evidence
of existence of new physics beyond the SM.
We see that the production cross sections of $H^\pm$ in the LR model
have lower bounds depending upon the charged Higgs mass.
The cross sections in the LR model are generically larger 
than those of the 2HD model in the most region of $\tan \beta$
and the better visibility of $H^\pm$ is expected
through $H^\pm \to t b$ channel.
We find that the decay into $\tau \bar{\nu}$ channel
is strongly suppressed in the LR model due to 
the smallness of $m_\tau^2/m_t^2$.
Combining the cross section and the ratio
$\Gamma(H^+ \to \bar{\tau} \nu)/\Gamma(H^+ \to t \bar{b})$,
therefore,
we can discriminate the LR model from the 2HD model
as the underlying physics in the charged Higgs sector.
If the production cross section of $H^\pm$ is lower 
than the minimal value predicted by the LR model
shown in Fig. 2,
we can conclude that the LR model is not the underlying physics.
When the $H^\pm$ boson is observed with the production cross section 
large enough,
it may be the ingrediant of either the LR model or the 2HD model
with large $\tan \beta$.
In that case,
the ratio $\Gamma(H^+ \to \bar{\tau} \nu)/\Gamma(H^+ \to t \bar{b})$
close to 0 indicates the LR model
and the ratio of 10 $\%$ indicates the 2HD model with large $\tan \beta$.

In conclusion, the charged Higgs sector can be crucially tested
in the production and decay at the LHC.
We study the LHC phenomenology of the charged Higgs boson in the LR model
compared with that in the 2HD model.

\acknowledgments
This work was supported by the Korea Research Foundation Grant 
funded by the Korean Government 
(MOEHRD, Basic Research Promotion Fund) 
(KRF-2007-C00145) 
and 
the BK21 program of Ministry of Education (K.Y.L.).

\def\PRD #1 #2 #3 {Phys. Rev. D {\bf#1},\ #2 (#3)}
\def\PRL #1 #2 #3 {Phys. Rev. Lett. {\bf#1},\ #2 (#3)}
\def\PLB #1 #2 #3 {Phys. Lett. B {\bf#1},\ #2 (#3)}
\def\NPB #1 #2 #3 {Nucl. Phys. {\bf B#1},\ #2 (#3)}
\def\ZPC #1 #2 #3 {Z. Phys. C {\bf#1},\ #2 (#3)}
\def\EPJ #1 #2 #3 {Euro. Phys. J. C {\bf#1},\ #2 (#3)}
\def\JHEP #1 #2 #3 {JHEP {\bf#1},\ #2 (#3)}
\def\JPG #1 #2 #3 {J. of Phys. G {\bf#1},\ #2 (#3)}
\def\IJMP #1 #2 #3 {Int. J. Mod. Phys. A {\bf#1},\ #2 (#3)}
\def\MPL #1 #2 #3 {Mod. Phys. Lett. A {\bf#1},\ #2 (#3)}
\def\PTP #1 #2 #3 {Prog. Theor. Phys. {\bf#1},\ #2 (#3)}
\def\PR #1 #2 #3 {Phys. Rep. {\bf#1},\ #2 (#3)}
\def\RMP #1 #2 #3 {Rev. Mod. Phys. {\bf#1},\ #2 (#3)}
\def\PRold #1 #2 #3 {Phys. Rev. {\bf#1},\ #2 (#3)}
\def\IBID #1 #2 #3 {{\it ibid.} {\bf#1},\ #2 (#3)}

\end{document}